\documentclass[english,prl,twocolumn,floatfix]{revtex4}

\usepackage[T1]{fontenc}
\usepackage[latin1]{inputenc}
\usepackage{amsmath}
\usepackage{amssymb}
\usepackage{babel}
\usepackage{epsfig}
\usepackage{pifont}

\def\vec#1{\mathbf{#1}}

\def\Zstar{Z^\star}
\def\kB{k_{\rm B}}
\def\lb{\lambda_{\rm B}}
\let\lB\lb
\def\cs{c_{\rm s}}
\def\Nabla{\boldsymbol\nabla}
\def\phic{\phi_{\rm c}}

\def\d#1{\!\mathrm{d}#1\,}

\begin{document}
\title{Nonlinear screening and gas-liquid separation in suspensions of charged colloids}

\author{Bas Zoetekouw}
\author{Ren\'e van Roij}

\affiliation{Institute for Theoretical Physics, Utrecht University,
Leuvenlaan 4, 3584CE Utrecht, the Netherlands}

\date{
    \today
}

\begin{abstract}
We calculate phase diagrams of charged colloidal spheres (valency
$Z$ and radius $a$) in a 1:1 electrolyte from multi-centered
nonlinear Poisson-Boltzmann theory. Our theory takes into account
charge-renormalization of the colloidal interactions and volume
terms due to many-body effects.  For valencies as small as $Z=1$
and as large as $10^4$ we find a gas--liquid spinodal instability
in the colloid-salt phase diagram provided $Z\lB/a\gtrsim24\pm1$,
where $\lB$ is the Bjerrum length.
\end{abstract}

\maketitle

Can like-charged colloids, dispersed in an aqueous solvent with
monovalent cations and anions, spontaneously demix into a
colloid-dilute ``gas'' phase and a colloid-dense ``liquid'' or
``crystal'' phase? For an index-matched solvent at room temperature,
the answer to this question is \emph{no} on the basis standard
linear screening theory as formulated in the 1940s by Derjaguin,
Landau, Verwey, and Overbeek (DLVO) \cite{dl1941,vo1948}, but
\emph{yes} on the basis of some intriguing experimental observations  in
quasi-deionized suspensions of charged colloids
\cite{ise1983,ito1994,grier,tataarora}.

The classic DLVO theory,
which is a corner stone of colloid science, predicts that pairs of
colloids of radius $a$ and charge $-Ze$ at separation $r$ interact
through a screened-Coulomb potential
\begin{equation} \label{v}
  v(r;Z,\kappa)
  =
  \kB T Z^2\lB
  \left(\frac{\exp[\kappa a]}{1+\kappa a}\right)^2
  \frac{\exp[-\kappa r]}{r},
\end{equation}
where the suspending medium is characterized by the temperature
$T$, the Bjerrum length $\lB=e^2/(\epsilon \kB T)$ with the
dielectric constant $\epsilon$, and the Debye screening length
$\kappa^{-1}$.  Here $\kB$ is the Boltzmann constant and $e$~the
proton charge.  The purely repulsive character of $v(r)$ does not
give rise to any cohesive energy to stabilize a dense liquid or
crystal phase in coexistence with a much more dilute gas phase.
Such cohesion effects, however, are observed in the experiments
of~Refs.\cite{ise1983,ito1994,grier,tataarora}. They include, for
instance, compressed crystal lattice spacings indicative of
gas--crystal coexistence \cite{ise1983}, gas-like voids in an
otherwise homogeneous liquid-like suspension \cite{ito1994},
long-lived metastable crystallites suggesting internal cohesion
\cite{grier}, and a (disputed) gas--liquid meniscus
\cite{tataarora}. Explaining any of these phenomena requires
cohesive energy, and the focus of much theoretical work has
therefore been on extending DLVO theory to find ``like-charge
attraction''.

Important ingredients beyond DLVO theory are ion-ion correlations,
nonlinear screening, and many-body effects. For (dilute) 1:1
electrolytes correlations are considered to be of only minor
importance and will be ignored here, whereas the other two are
important here. Nonlinear screening has been studied extensively
in the context of (spherical) cell models, where the geometry
allows for straightforward numerical solutions of the nonlinear
Poisson--Boltzmann (PB) equation
\cite{alexander,trizac-2003,levin2003,gruenberg-vanroij}. An
important nonlinear effect is quasi-condensation of ions onto a
highly charged colloidal surface when $Z\lB/a\gtrsim5-10$.  As a
consequence, the colloidal charge that appears in the prefactor of
Eq.(\ref{v}) is renormalized from its bare value $Z$ to a state
point dependent $\Zstar<Z$, i.e., the interactions are reduced but
remain repulsive. Interestingly, free-energy calculations on the
basis of the nonlinear cell model show \emph{no} sign of a
gas--liquid spinodal instability \cite{gruenberg-vanroij} either.
On the basis of these results, together with, e.g., formal proofs
\cite{neu+trizac} that nonlinear PB theory yields purely repulsive
pair interactions, it is tempting to conclude that gas-liquid
coexistence is impossible within (non)linear screening theory.

However, there are also studies that do show cohesion and
gas--liquid spinodal instabilities. Examples include the early
work by Langmuir \cite{langmuir}, primitive model simulations of
asymmetric electrolytes \cite{linse,panag,antti}, PB calculations
showing triplet attractions on top of pairwise repulsions
\cite{russ-PRE} in agreement with experiments
\cite{russ-EPL+JPCM}, and the extended Debye-H\"uckel theory and
the boot-strap PB theory of
Refs.\cite{chanlinsepetris,petrischanlinse}. Interestingly, these
systems have the explicit colloidal many-body character in common,
as opposed to the cell geometry discussed above. Unfortunately, it
is extremely time consuming and practically impossible to extend
simulations such as those of Refs.\cite{linse,panag,antti} to
realistic colloidal parameters (say, $Z=10^3$--$10^4$ with finite
salt concentrations), or to calculate or measure effective
$n$-body potentials with $n\ge4$. Also attempts to study the full
nonlinear many-body system, e.g., within the schemes as proposed
in Refs.\cite{fushiki,dobnikar,loewen-hansen-madden}, turn out to
be computationally difficult in the colloidal parameter regime. In
fact, the colloidal parameter regime has so far mainly been
studied as an explicit many-body system within \emph{linear}
screening theory, where the many-body character appears as a
nontrivial density-dependence of pair interactions and as volume
terms that can drive a gas--liquid transition
\cite{beresford,warren,denton,diehl,bas_grcan}, albeit mostly in
regimes where charge renormalization should have been taken into
account thereby stabilizing the suspension \cite{diehl}.

The key question addressed in this Letter is whether the intriguing and
sometimes hotly-debated experiments like those of
Refs.\cite{ise1983,ito1994,grier,tataarora} can be
explained by hard-core repulsions and Coulomb interactions only. This
question is answered by calculating phase diagrams of the primitive model in
the colloidal limit ($Z\gg 1$ and point-like cations and anions). We retain
the best aspects of two well-established theories by combining  the
nonlinear screening effects of cell theory with the explicit many-body
character of linear screening theory.

We consider $N$ spherical colloids of hard-core radius $a$ and
fixed charge $-Ze$ at positions~$\{\vec R_i\}$ in a solvent of
dielectric constant $\epsilon$ and volume $V$ at temperature $T$.
The system is in osmotic contact with a salt reservoir of
monovalent point-like cations and anions at pair density $2\cs$,
which gives rise to (yet unknown) ion concentrations $c_{\pm}$ in
the suspension. We only consider pairwise Coulomb and hard-core
potentials between any pair of particles, such that contact
distances are $2a$, $a$, and $0$ for a colloid--colloid,
colloid--ion, and ion--ion pair, respectively.  Thermodynamic
properties and the phase diagram of this system can be determined
once we explicitly know the semi-grand potential $F(N,V,T,\cs)$,
which describes the colloids canonically and the ions
grand-canonically, and which is defined by the partition function
$\exp[-\beta F]=(1/N!)\int_V \d{\vec R_1}\cdots \d{\vec
R_N}\exp[-\beta H\{\vec R\}]$. Here $\beta=1/(\kB T)$ and
$H(\{{\vec R}\})$ is the effective colloid Hamiltonian. It was
shown in Ref.\cite{bas_grcan} that $H$ is the sum of the bare
colloid--colloid interactions (hard-core and unscreened Coulomb)
and the grand potential of the inhomogeneous fluid of cations and
anions in the external field of the fixed colloidal charge
distribution~$q({\vec r})=-Z\sum_{i=1}^N\delta(|{\vec r}-{\vec
R}_i|-a)/4\pi a^2$. Within mean-field theory, the effective
Hamiltonian can be written as a sum of entropic and
electrostatic-energy terms \cite{bas_grcan}
\begin{align}\begin{split}
\beta H = &\beta H_{\mathrm{HS}}+
  \sum_{\alpha=\pm} \int\d{\vec r} \rho_{\alpha}(\vec r)
    \left[ \ln\frac{\rho_\alpha(\vec r)}{\cs} - 1
    \right] \\
&+ \frac{1}{2}\int\d{\vec r}
    [\rho_+(\vec r)-\rho_-(\vec r)+q(\vec r)]\phi(\vec r) ,
\label{H}
\end{split}\end{align}
where $H_{\mathrm{HS}}$ denotes the hard-core interactions between
the colloids, and where all ${\vec r}$-integrals are over the
volume outside the colloidal cores: $|{\vec r}-{\vec R}_i|\geq a$
for all $i=1,\dots,N$.  The yet unknown quantities are the
equilibrium density profiles $\rho_\pm({\vec
r})=\cs\exp[\mp\phi({\vec r})]$ of the cations and the anions, and
the (dimensionless) electric potential~$\phi({\vec r})$.  Note
that $\phi(\vec r)\equiv0$ in the reservoir. These unknowns follow
from Poisson's equation $\Nabla^2\phi({\vec
r})=-4\pi\lB[\rho_+(\vec r)-\rho_-(\vec r)+q({\vec r})]$, which
can be recast as the multi-centered non-linear PB equation
\begin{subequations}
\begin{align}
  \Nabla^{2}\phi(\vec{r}) &=
    \kappa^{2}\sinh\phi(\vec{r}) &&
    \text{$\vec r$ outside cores}\label{PBE}\\
  {\vec n}_i\cdot\Nabla\phi({\vec r}) &=
    Z\lB/a^2 &&
    {\vec r}={\vec R}_i+a{\vec n}_i,\label{bc}
\end{align}
\end{subequations}
where~$\kappa^{-1}=1/\sqrt{8\pi\lb\cs}$ is the reservoir's Debye
screening length, and where ${\vec n}_i$ is the unit surface
normal of colloid $i=1,\dots N$.

In order to evaluate~$H$ of Eq.\eqref{H}, we approximately solve
Eqs.\eqref{PBE}--\eqref{bc} as follows.  We imagine each colloid
$i=1,\dots,N$ in the center of a virtual cell of yet unknown
radius~$b\geq a$, and assume that the potential inside each cell
is spherically symmetric and given by $\phic(|{\vec r}-{\vec
R}_i|)$ for $a<|\vec r-\vec R_i|<b$. Denoting the net
(yet-unknown) charge of the cell by $Q$, the cell-potential
$\phic(r)$ for $a<r<b$ is the solution of the radially symmetric
PB equation
\begin{align}\label{PB-cell}
\begin{split}
& \frac{1}{r^2}\frac{\mathrm{d}}{\mathrm{d}r}
    \left[r^2\frac{\mathrm{d}}{\mathrm{d}r} \phic(r)\right]
  = \kappa^2\sinh\phic(r); \\
& \phic'(a) = \frac{Z\lb}{a^2}; \qquad
  \phic'(b) = \frac{Q\lb}{b^2}.
\end{split}
\end{align}
The boundary value problem~\eqref{PB-cell} is easy to solve
numerically on a radial grid for given $\kappa$, $b$, $\lB$, $Z$
and~$Q$. Outside the cells we retain the multi-centered character
of $\phi({\vec r})$, but we exploit that it varies only weakly from
some spatial constant $\bar\phi$ (provided $b$ is large enough),
such that $\phi({\vec r})-\bar\phi$ is the small parameter in a
linearized treatment of Eq.\eqref{PBE}. The linearized
multi-centered PB equation can be solved analytically
\cite{bas_grcan}, and in terms of a (yet unknown) effective
colloidal charge $-\Zstar e$ the resulting potential outside the cells
is given by  $\phi({\vec r})=\bar\phi -\tanh\bar\phi
+\sum_{i=1}^{N}\phi_1(|\vec r-\vec R_i|)$, with the Yukawa
``orbitals''
\begin{align}\label{linprof}
  \phi_1(r) =
    -\Zstar\lb
    \frac{\exp(\bar\kappa a)}{1+\bar\kappa a}
    \frac{\exp(-\bar\kappa r)}{r}.
\end{align}
Following Ref.\cite{bas_grcan}, we identify~$\bar\phi$ with the
Donnan potential of the renormalized system, such that
$\sinh\bar\phi=-\Zstar n/(2\cs\exp[-\eta/(1-\eta)])$, with $n=N/V$
the colloid density and $\eta=4\pi{a^3}n/3$ the packing fraction.
Hence also $c_\pm=\cs\exp[\mp\bar\phi]$ and
$\bar\kappa^2=4\pi\lB(c_++c_-)$ are known explicitly.  Therefore,
$\phi(\vec r)$ is completely determined inside and outside the
cells once $Q$, $b$ and $\Zstar$ are specified.

How to determine these unknown quantities? In this Letter we
calculate $Q$ and $Z^*$, for a fixed $b$ to be discussed below, by
imposing continuity of the (average) potential and its gradient at
the cell boundaries. Choosing the origin at ${\vec R}_1$, this
implies that
\begin{subequations}
\begin{align}
  \label{eq:pot_cont1}
  \phic(b) &= \bar{\phi} - \tanh\bar{\phi} + \phi_1(b)
    + \Bigl\langle
      \sum_{i=2}^{N}\phi_1(|b\vec n_1 - \vec R_i|)
    \Bigr\rangle;
  \\
  \label{eq:E_cont1}
  \phic'(b) &= \phi_1'(b)
    + \Bigl\langle
      \vec n_1 \cdot \Nabla
      \sum_{i=2}^{N}\phi_1(|b\vec n_1-\vec R_i|)
    \Bigr\rangle.
\end{align}
\end{subequations}
The angular brackets in the right hand side of
Eq.\eqref{eq:pot_cont1} indicate an average involving the
colloid--colloid radial distribution function $g(R)$, and this
term can be written as $n\int\d{\vec R}g(R)\phi_1(|b\hat{\vec
r}-{\vec R}|)$, and likewise for Eq.\eqref{eq:E_cont1}. In
principle one could think of setting up a scheme to calculate
$g(R)$ and the effective hamiltonian $H(\{{\bf R}\})$
self-consistently, yet here we are satisfied with the crude
approximation that $g(R)=0$ for $R<2b$ and 1 for $R>2b$. This
simplification allows for straightforward analytic expressions for
the bracketed terms in Eqs.(\ref{eq:pot_cont1}) and
(\ref{eq:E_cont1}), which can then be easily solved numerically
for the two unknowns $Q$ and $Z^*$ at fixed ($Z, \lB, a, b,\eta$).

The remaining problem is to choose the cell radius $b$. We checked
that the largest physically reasonable choice, $b=a\eta^{-1/3}$,
for which the cell has the volume of the Wigner-Seitz cell, yields
essentially charge-neutral cells, $Q/Z\ll 1$, such that (i)
$\Zstar$
is identical to Alexander's renormalized charge
\cite{alexander,trizac-2003}, and (ii) {\em no} gas-liquid
instability is expected \cite{gruenberg-vanroij}. One also easily
checks that the smallest possible choice, $b=a$, ignores any
nonlinearity and leads essentially to the volume-term theories of
Refs.\cite{warren,denton,bas_grcan}, which {\em do} predict
gas-liquid spinodals. Our present choice for $b$ is in between
these two extremes, and is such that $b$ is large enough for the
region outside the cells to be properly described by linear
screening, yet small enough to avoid significant overlaps of the
cells. Moreover, since $b$ is an unphysical parameter, the phase
diagrams should not depend on $b$. All this is achieved, except
perhaps in some extreme parameter regimes, by choosing $b$ such
that $|\phi_c(b)-\bar\phi|=\delta$ (or $b=a$ if $Z\lB/a$ is small
enough), with fixed $\delta\simeq 1$. This criterion for $b$ leads
to values of $b/a$ changing monotonically from~$b/a\rightarrow1$
(linear screening) for high packing fractions ($\eta>0.5$) or high
salt concentrations ($\kappa a>5$) to $b/a>10$ (nonlinear
screening) for dilute systems ($\eta\ll10^{-3}$) at extremely low
salt concentrations ($\kappa{a}<0.1$).

Our numerical results show  that $\Zstar$ as defined by our
procedure has all the characteristics of the renormalized charge
as defined by Alexander\cite{alexander}, i.e. $\Zstar\simeq Z$ if
$Z\lB/a\lesssim 5$, if $\kappa a$ is sufficiently large, or if
$\eta$ is high, and $\Zstar<Z$ otherwise. For instance, for
$Z=1000$ and $a=100\lambda_B$ (which are of the order of the
experiments of Refs.\cite{tata1997}), we find for $\eta<10^{-2}$
that $\Zstar\simeq 700,900$ for $\kappa a=1,5$, respectively, and
$\Zstar$ increases (in a non-monotonic fashion) to $Z$ for
increasing $\eta$. These effects, which are independent of
$\delta$ (and hence of $b$) provided $0.75\leq\delta\leq1.25$,
are also found in Alexander's traditional cell model
\cite{belloni} and in experiments \cite{shapran}.

With $Q$ and $\Zstar$ determined from the continuity at the cell
boundary \eqref{eq:pot_cont1} and~\eqref{eq:E_cont1}, and $b$ from
the criterion as described above, the potential $\phi({\vec r})$,
and hence the ionic density profiles $\rho_\pm({\bf r})$, are
known explicitly, both inside and outside the cells. The
Hamiltonian~\eqref{H} can thus be evaluated explicitly. After
tedious but straightforward calculation, in which the logarithmic
terms are expanded to quadratic order outside the cells, one
obtains
\begin{equation}\label{eq:H}
  H(\{{\vec R}\})
  =
  \Phi
  +
  \sum_{i<j}^N
  v(|\vec R_i-\vec R_j|;
  \Zstar\sqrt{1-\Upsilon(\bar\kappa a)},
  \bar\kappa)
\end{equation}
with a so-called volume term $\Phi$ that does \emph{not} depend on
the coordinates of the colloids, and a sum of pairwise
screened-Coulomb interactions~\eqref{v} with an effective charge
and an effective screening parameter.  Here we introduced the
function~$\Upsilon(x)=(1+x)(1-\exp[-2x])/2x$, such that the factor
$\sqrt{1-\Upsilon(\bar\kappa{a})}$ in the effective charge is of
order $1/2$ in all but very extreme parameter regimes. The volume
term $\Phi$ in Eq.\eqref{eq:H} is given by
\begin{align}\label{eq:Phi}
  &\nonumber
  \frac{\beta\Phi}{V} =
  4\pi\cs{n} \int_a^b\d{r} r^2\left\{
    \phic(r)\sinh\phic(r) - 2\cosh\phic(r)
  \right\}
  \\&\null~
  -\frac{{n}}{2}Z\phic(a)
  +\frac{\eta}{1-\eta}\frac{2c_+c_-}{c_++c_-}
  \\&\null~\nonumber
  +\left[1-\eta\frac{b^3-a^3}{a^3}\right]
    \left\{
      \frac{n}{2} \Zstar \bar\phi
      + \sum_{\alpha=\pm} c_\alpha
        \left[\ln\frac{c_\alpha}{\cs} -1\right]
    \right\}.
\end{align}
One easily checks that in the limit $b\rightarrow a$, for which
$\Zstar\rightarrow{Z}$ and
$\phic(a)\rightarrow\bar\phi-\tanh\bar\phi+\phi_1(a)+
\sum_{i=2}^{N}\phi_1(|a\vec n_1-\vec R_i|)$, one recovers
essentially the linear screening theory of Ref.\cite{bas_grcan},
and that the first term of the second line of Eq.(\ref{eq:Phi})
contains the Debye-H\"{u}ckel like $-n^{3/2}$ cohesive energy (per
unit volume) that drives the spinodal gas-liquid instability at
low salinity \cite{bas_grcan,warren,denton}.

\begin{figure}
\epsfig{file=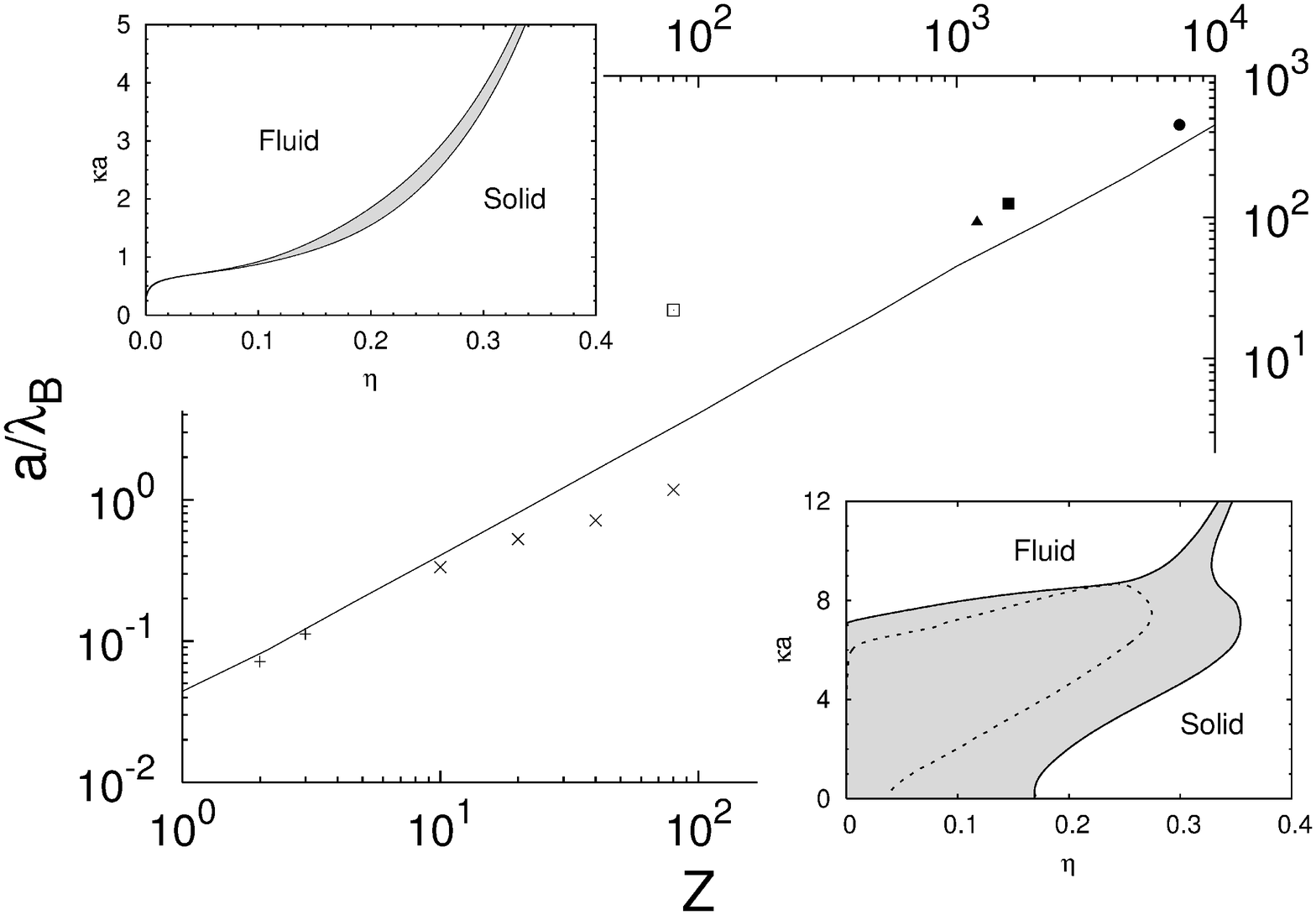,width=\hsize}
    \caption{
  Typical $\eta-\kappa a$ phase diagrams (insets) and critical
  line $Z\lB/a\sim24\pm 1$ (main figure), as determined over many
  decades of the parameters by the present theory, for suspensions
  of charged colloids (charge $-Ze$, radius $a$, packing fraction
  $\eta$) in osmotic contact with a $1$:$1$ electrolyte of reservoir
  screening constant $\kappa$ and Bjerrum length $\lambda_B$. The
  critical line, which actually consists of three superimposed
  lines for $\delta=0.75,1,1.25$ (see text), separates the
  strong-coupling regime (lower right) \emph{with} a spinodal
  gas--liquid instability (dashed line) and a large density gap
  (grey area) at fluid--solid coexistence for low $\kappa a$ from
  the low-coupling regime (upper left part) \emph{without}
  spinodal instability and with only a narrow density gap at
  fluid--solid coexistence. Criticality as found in the primitive
  model simulations of Refs.\cite{linse} and~\cite{panag} are
  indicated by $\times$ and $+$, respectively, the simulated state
  point of Ref.\cite{antti2} without an instability is indicated
  by $\square$, and the experimental systems in which evidence for
  large density gaps at gas-liquid and gas-crystal coexistence was
  found are indicated by \ding{115}~\cite{monovoukas},
  \ding{110}~\cite{tata1997}, and \ding{108}~\cite{grier}.
}
\label{fig:meta}
\end{figure}

With the Hamiltonian explicitly given by Eq.\eqref{eq:H}, we can
calculate phase diagrams from $F=\Phi+F_{\mathrm{HSY}}$, with
$F_{\mathrm{HSY}}$ the free energy of a hard-sphere Yukawa fluid,
which we calculate as in Ref.\cite{bas_grcan} by exploiting the
Gibbs--Bogoliubov inequality.  For fixed $a/\lB$ and $Z$, we
calculate phase diagrams in the $\eta$--$\kappa{a}$
representations by imposing equal osmotic pressure and chemical
potential in the coexisting phases.  The insets of
Fig.\ref{fig:meta} show two typical phase diagrams, calculated
with $\delta=1$, where the upper left corner ($Z=10$, $a/\lB=100$)
only shows crystallization with a narrow density gap, and the
opposite corner ($Z=2000$, $a/\lB=10$) exhibits a spinodal
instability at low $\kappa a$ and hence a large density gap
between the coexisting phases. From many such phase diagrams we
constructed a curve in the ($Z$,$a/\lB$) plane of
Fig.\ref{fig:meta}, below which a spinodal instability is present
in the $\eta$--$\kappa{a}$ plane.  For $\delta=0.75$, $1$, and
$1.25$, these curves superimpose on a single line that is well
approximated by $Z\lB/a=24\pm1$ over many decades of
$(Z,a/\lambda_B)$. The independence of $\delta$ confirms again
that the phase diagrams are independent of the artificial cell
radius $b$ (provided $b$ is chosen sensibly, i.e. ${\cal
O}(\delta)=1$). The critical line from the linear screening
theory, $b=a$, satisfies $Z\lambda_B/a\simeq 6-7$ (not shown) for
$Z\geq 10$, where the weaker critical coupling indicates that
charge renormalization has a stabilizing effect on the suspension
in agreement with Ref.\cite{diehl}.

Fig.\ref{fig:meta} also shows the parameters of three colloidal
systems for which phase-instabilities have been observed
experimentally (filled symbols in the figure); they are reasonably
close to, yet systematically above, our predicted critical line,
by a factor of $1.5$--$2$. Our critical line also shows good
agreement with (estimates of) critical points in the salt-free
primitive model simulations of Refs.~\cite{linse,panag} for
$10\leq Z\leq 80$ ($\times$) and $Z=2,3$ ($+$), respectively,
although a systematically increasing deviation up to a factor
$\sim3$ (for $Z=80$) is obscured by the double logarithmic scale
of Fig.1. This deviation could perhaps be explained by the fact
that for close-to-critical values of $Z\lB/a$, instabilities
(detached from the freezing line) first appear at $\kappa a=0$ but
only for extremely dilute systems ($\eta\ll10^{-3}$, decreasing
with increasing $Z$), while the lowest density considered in
Ref.\cite{linse} is as "high" as $\eta=0.00125$; the instability
reaches this state point at a coupling that increases with $Z$
\cite{bas_lang}. Note also that for $Z=1$, our theory predicts the
critical point at $a/\lB\approx0.048$, which is close to the
well-known critical point of the symmetric $1$:$1$ electrolyte at
$a/\lB=0.057$ \cite{mcgahay+fisher}. It is perhaps surprising, yet
comforting, that the critical point as predicted by the present
theory, which is "designed" to deal with $Z\gg 1$, agrees
quantitatively with the primitive model simulations of in the
low-valency regime $Z\leq 10$.

To conclude, we have constructed a theory for colloidal
suspensions that interpolates between the well known limits of
linear DLVO-type and non-linear cell-type PB theories, thereby
combining the multi-centered character and the volume terms of the
former with the charge renormalization of the latter.  For high
enough charges and small enough radii ($Z\lB/a\gtrsim24\pm1$, i.e.
well into the charge renormalization regime), we find spinodal
instabilities at low ionic strengths. The theory directly extends
the undisputed gas--liquid instability for  asymmetric low-valency
electrolytes ($Z\leq 10$) to the colloidal regime ($Z\gtrsim
1000$), although the existing experimental evidence in the
colloidal regime is at somewhat weaker couplings than required
according to our predictions here. We are currently extending the
present theory to take charge regulation into account, in order to
investigate whether the required coupling can be shifted towards
experimentally realized systems. Alternatively it is interesting
to consider the possibility that the experimentally determined
``effective'' colloid charge should not be identified with $Z$ (as
we did here) but with a renormalized charge \cite{palberg}, as
this would shift the experimental points of Fig.\ref{fig:meta}
closer to or beyond the critical line predicted here.

We thank Marjolein Dijkstra for useful discussions. This work is
part of the research programme of the ``Stichting voor
Fundamenteel Onderzoek der Materie (FOM)'', which is financially
supported by the ``Nederlandse Organisatie voor Wetenschappelijk
Onderzoek (NWO)''.


\bibliographystyle{myapsrev}

\end{document}